\documentclass{article}
\usepackage{amssymb}
\usepackage{amsmath}
\usepackage{graphicx}

\newtheorem{theorem}{Theorem}
\newtheorem{acknowledgement}[theorem]{Acknowledgement}

\input{tcilatex}
\begin{document}

\title{Diffusion of two brands in competition: cross-brand effect}
\author{C. E. Laciana$^{\ast \dagger }$, G. Gual$^{\ast }$, D. Kalmus$^{\ast
}$, N. Oteiza-Aguirre* \and and S. L. Rovere* \\
\\
$^{\ast }$Grupo de Aplicaciones de Modelos de Agentes (GAMA), \\
Facultad\ \ de Ingenier\'{\i}a, Universidad de Buenos Aires, \ \\
Avenida Las Heras 2214 Ciudad Aut\'{o}noma de Buenos\\
Aires, C1127AAR, Argentina. \\
clacian@fi.uba.ar \\
$^{\dagger }$Instituto de Investigaci\'{o}n en Ciencias Econ\'{o}micas de \\
la Universidad del Salvador (USAL), calle Marcelo T. \\
de Alvear 1335, Ciudad Aut\'{o}noma de Buenos Aires, \\
C1058AAU, Argentina. 
Cross-brand effect; Potts model; \\
Equilibrium points; Models comparison }
\date{ }
\maketitle

\begin{abstract}
We study the equilibrium points of a system of equations corresponding to a
Bass based model that describes the diffusion of two brands in competition.
To increase the understanding of the effects of the cross-brand parameters,
we perform a sensitivity analysis. Finally, we show a comparison with an
agent-based model inspired in the Potts model. Conclusions include that both
models give the same diffusion curves only when the cross coeficients are
not null.
\end{abstract}

\section{\protect\bigskip Introduction}

It is of present interest, from an economic point of view, to fully
understand the processes and drivers behind the diffusion of innovations.
The existence of many recent papers reviewing this subject, such as \cite%
{Mahajan}, \cite{Meade} and \cite{Peres} are an example of its relevance. A
pioneer work is the well-known Bass model \cite{Bass 69} which describes the
curves of adoption for many durable goods with great precision.

Keeping in mind the success of the Bass model in the description of the
diffusion process of many new products, it is natural to extend the
formalism to describe the adoption curves of two brands which are launched
simultaneously and dispute the same market. This possibility has been
previously investigated, as we can see in  \cite{Libai} and \cite{Peres}.

We have, within the Bass formalism, two coupled differential equations. The
simplest coupling is the one associated to the competition between two
brands within a common market. However, as in \cite{Savin}, a cross-brand
effect can be introduced that takes into account the interactions between
the adopters of one brand with the potential adopters of the other brand.
This adds new coupling parameters in the equation that further describe the
dynamics of product adoption, these will be referred to as cross-brand
parameters from here on.

The problem of two brands in competition can also be approached from a
microscopic point of view, where both the preferences of each individual and
its interaction with others are taken into account. In a previous work \cite%
{Laciana et al.} it was shown, for an innovation diffusing in a market, that
it is possible to relate the microscopic variables of an agent-based model
(ABM), to the parameters of the Bass model. For that purpose, a physical
analogy was used, namely, the well-known statistical Ising model was adapted
for the study of technology diffusion \cite{Laciana-Rovere}.

Later on, a generalization for many options was described \cite%
{Laciana-Oteiza}, using an analogy with the statistical Potts model \cite%
{Potts} which lets us consider "n" options in competition. There are other
approaches for modeling competition between brands, such as the case of ref. 
\cite{Peres} cited before. Also, in ref. \cite{Sengupta} the problem of
competition between two brands is studied in terms of: "innovate or copy".
There, the algorithm comes from the Logit model which, because of the
threshold condition used, is analogous to our Potts model with
\textquotedblleft zero temperature\textquotedblright . Another example of an
agent based model for competition between options is the one used in ref. 
\cite{Casillas}. In that paper the Monte Carlo method is used and companies
(agents), can choose between two options (products or services) by means of
a mechanism based on costs and payoffs.

One of the goals of the present work is to relate the microscopic variables
of the agent-based model to the macroscopic variables of the Bass based
model for the case of two brands in competition. In particular, we focus our
research in a systematic study of the cross-brand terms, looking for the
values that fit the ABM for two brands. We do this with the hypothesis that,
except for the macroscopic parameters associated to the cross-brand effect,
all the other parameters can be equal to the ones corresponding to the
isolated brands (without competition). Although the best fit between the
microscopic and macroscopic models is achieved by varying all of the
microscopic variables, this hypothesis can be considered approximately
valid, which encourages the generalization of the line traced in ref. \cite%
{Laciana-Rovere} regarding the process of correspondence between the
macroscopic and microscopic models.

This paper is organized as follows: In Section 2, we introduce the system of
two differential equations that describes the dynamics of two brands in
competition, we perform the analysis of the equilibrium points of the system
and show the influence of the value of the parameters on those points. In
Section 3, we provide the n-optional formalism used in the ABM. In Section
4, we perform the comparison between the dynamics emerging from the two
considered models (i.e. Bass-like and ABM). In Section 5, we summarize the
main conclusions.

\section{Coupled Bass System}

In the original Bass article \cite{Bass 69} the aggregate adoption rate of a
new product (consumer durable good), in a given potential market ($m$) is
calculated as a function of two kinds of parameters, each describing two
different types of influence: the innovation parameter ($p$) reflects
people's intrinsic tendency to adopt an innovation, while the imitation
parameter ($q$) reflects the "word of mouth" or the "social contagion",
representing the positive influence that people that has already adopted
makes on potential adopters.

Such as stressed in \cite{Libai}, when two brands are considered, we can
identify two kinds of effects related to the interaction between adopters
and potential adopters: one is known as "within-brand" and the other as
"cross-brand". The first one is the influence of the adopters of a brand on
the probability that potential adopters will adopt the same brand. The
second is the positive effect produced by adopters of a brand on the
probability that potential adopters will adopt the other brand. Libai
observed the cross-brand effect in Apple's launch of the iPhone in 2007
where word-of-mouth trasmission of the product's particularities
incentivated the sales not only of the iPhone but of the whole smartphone
category.

Using the same equations than ref. \cite{Libai} we have

\begin{equation}
\frac{dN_{1}}{dt}\left( t\right) =\left[ p_{1}+q_{11}\frac{N_{1}\left(
t\right) }{m}+q_{12}\frac{N_{2}\left( t\right) }{m}\right] \left( m-N\right)
,  \label{1}
\end{equation}

\begin{equation}
\frac{dN_{2}}{dt}\left( t\right) =\left[ p_{2}+q_{22}\frac{N_{2}\left(
t\right) }{m}+q_{21}\frac{N_{1}\left( t\right) }{m}\right] \left( m-N\right)
.  \label{2}
\end{equation}

Where $N_{1}$ and $N_{2}$ are respectively the number of adopters of brands
1 and 2 , $m$ is the common potential market, $N\left( t\right) =N_{1}\left(
t\right) +N_{2}\left( t\right) $ is the total number of adopters at time $t$%
, $p_{1}$ and $p_{2}$ are the external influence parameters for brands 1 and
2 respectively, $q_{11}$ and $q_{22}$ are the within-brand influence
parameters for brand 1 and brand 2 respectively, $q_{12}$ is the cross-brand
influence of brand 2 on brand 1 and conversely $q_{21}$ of brand 1 on 2.

Such as indicated in ref. \cite{Libai} there is some bibliography related
with the last formulation where the approach $q_{ii}=$ $q_{ij}=q_{ji}$ or
even $q_{ij}=0$ is used. However, those coefficients are never considered to
be negative. A negative value of the cross-brand term would mean that the
consumer that adopted brand 2 for example, would generate a positive
influence to adopt brand 2 on potential adopters, but this would be
redundant, since that effect is already considered in Eq. (\ref{2}) through $%
q_{22}$. On the other hand, a positive value por the cross-brand term would
mean a reinforcement for the purchase of brand 1 by adopters of brand 2 and
this would make sense if these adopters have had a bad experience with said
brand. Therefore the only restriction that we employ on the cross
coefficients is $q_{ij}\geqslant 0$.

In the next subsection we will analyze the equilibrium points reached by the
system of equations 1 and 2 and its dependence with the values of the
parameters.

\subsection{Analysis of equilibrium points}

In order to obtain a solution that is independent of the potential market
(assumed constant by the model), it's convenient to use the dimensionless
formulation. We then have the following:

\begin{equation}
\overset{\cdot }{n_{1}}\left( t\right) =\left(
p_{1}+q_{11}n_{1}+q_{12}n_{2}\right) \left( 1-n_{1}-n_{2}\right) ,  \label{3}
\end{equation}

\begin{equation}
\overset{\cdot }{n_{2}}\left( t\right) =\left(
p_{2}+q_{22}n_{2}+q_{21}n_{1}\right) \left( 1-n_{1}-n_{2}\right) ,  \label{4}
\end{equation}

with $\overset{\cdot }{n_{i}}\equiv dn_{i}/dt,$ $n_{i}\equiv N_{i}/m$ and $%
i=1,2$.

The equilibrium points are those that satisfy $\overset{\cdot }{n_{1}}=%
\overset{\cdot }{n_{2}}=0.$ They will be the ones that make one (or both) of
the parentheses that compose the equations (\ref{3}) and (\ref{4}) null.
However, since all the parameters must be positive to preserve their
conceptual meaning, the first parentheses cannot be null. Therefore, the
reasonable equilibrium condition will be

\begin{equation}
1-n_{1}-n_{2}=0.  \label{5}
\end{equation}

Eq. (\ref{5}) represents all the pair of values $\left( n_{1},n_{2}\right) $
for which the market is saturated. They can be expressed \ as $\left(
n_{1},n_{2}\right) =\left( n_{1},1-n_{1}\right) $ with $n_{1}>0,$ $n_{2}>0$.

We will now proceed with a sensitivity analysis of these equilibrium points.
In order to do so, we perform an infinitesimal displacement given by

\begin{equation}
n_{1}\longrightarrow n_{1}+\delta n_{1},\text{ \ }n_{2}\longrightarrow
n_{2}+\delta n_{2}  \label{6}
\end{equation}

Replacing the perturbed quantities (\ref{6}) in the equations (\ref{3}) and (%
\ref{4}), neglecting quadratic terms and eliminating $\overset{\cdot }{n_{1}}
$ and $\overset{\cdot }{n_{2}}$ by means of the substitution of the
equations (\ref{3}) and (\ref{4}), regrouping and replacing the Eq.(\ref{5}%
), we finally obtain

\begin{equation}
\overset{\cdot }{\delta n_{1}}\simeq \left[ n_{1}\left( q_{12}-q_{11}\right)
-\left( p_{1}+q_{12}\right) \right] \left( \delta n_{1}+\delta n_{2}\right) ,
\label{7}
\end{equation}

\begin{equation}
\overset{\cdot }{\delta n_{2}}\simeq \left[ n_{1}\left( q_{22}-q_{21}\right)
-\left( p_{2}+q_{22}\right) \right] \left( \delta n_{1}+\delta n_{2}\right) .
\label{8}
\end{equation}

Moreover, the following must also be satisfied

\begin{equation}
n_{1}+\delta n_{1}+n_{2}+\delta n_{2}\leqslant 1.  \label{9}
\end{equation}

Eq. (\ref{9}) shows that two possibilities exist for any given time: that
the market is saturated or that it didn't still reach the saturation. Due to
the fact that $n_{1}$ and $n_{2}$ satisfy Eq. (\ref{5}) the perturbations
satisfy the inequality

\begin{equation}
\delta n_{1}+\delta n_{2}\leqslant 0.  \label{10}
\end{equation}

We will now analyze the two situations implied in the inequality (\ref{10}):

i) When the equality is valid, in agreement with Eq. (\ref{9}), the
displacements are on a straight line, then

\begin{equation}
\delta n_{1}=-\delta n_{2}.  \label{11}
\end{equation}

Replacing Eq. (\ref{11}) in Eqs. ( \ref{7}) and (\ref{8}) we get

\begin{equation}
\overset{\cdot }{\delta n_{1}}\simeq 0,\text{ }\overset{\cdot }{\delta n_{2}}%
\simeq 0.  \label{12}
\end{equation}

Eq. (\ref{12}) tells us that the displacements remain constant in time,
which means that if we move a point on the straight line, the new position
is invariable along time. Therefore we can affirm that all the points on the
line are indifferent equilibrium points.

ii) For the situation where Eq. (\ref{10}) is strictly negative
(inequality), it is necessary to study the system of equations (\ref{7}) and
(\ref{8}) in order to determine if the displacement is an increasing
function with time or not. For that analysis it's convenient to represent
the system of equations in the following matrix form

\begin{equation}
\overset{\cdot }{\delta \mathbf{n}}\mathbf{=}\text{ }\mathbf{A}\delta 
\mathbf{n},  \label{13}
\end{equation}

with $\delta \mathbf{n\equiv }\left( 
\begin{array}{c}
\delta n_{1} \\ 
\delta n_{2}%
\end{array}%
\right) $ and $\mathbf{A}$ a matrix with the form:

\begin{equation}
\mathbf{A}\equiv \left( 
\begin{array}{cc}
a & a \\ 
b & b%
\end{array}%
\right) ,  \label{14}
\end{equation}

with

\begin{equation}
a\text{ }\mathbf{\equiv }\text{ }-\left( p_{1}+q_{12}\right) +\left(
q_{12}-q_{11}\right) n_{1}  \label{15}
\end{equation}

and

\begin{equation}
b\text{ }\mathbf{\equiv }\text{ }-\left( p_{2}+q_{22}\right) +\left(
q_{22}-q_{21}\right) n_{1}.  \label{16}
\end{equation}

To decouple the system of equations implicit in the vectorial equation (\ref%
{13}), we proceed to diagonalize matrix A, finding the most suitable change
of coordinates, that is:

\begin{equation}
\mathbf{A}=\mathbf{PDP}^{-1},  \label{17}
\end{equation}

with $\mathbf{D=}\left( 
\begin{array}{cc}
0 & 0 \\ 
0 & a+b%
\end{array}%
\right) ,$ $\mathbf{P=}\left( 
\begin{array}{cc}
1 & a/b \\ 
-1 & 1%
\end{array}%
\right) $ and $\mathbf{P}^{-1}=\frac{1}{1+a/b}\left( 
\begin{array}{cc}
1 & -a/b \\ 
1 & 1%
\end{array}%
\right) .$

Substituting Eq. (\ref{17}) in Eq. (\ref{13}) and multiplying by $\mathbf{P}%
^{-1}$ we have

\begin{equation}
\mathbf{DP}^{-1}\delta \mathbf{n=P}^{-1}\overset{\cdot }{\delta \mathbf{n}}.
\label{18}
\end{equation}

The above equation suggests the change of variables

\begin{equation}
\mathbf{P}^{-1}\delta \mathbf{n=}\text{ }\delta \mathbf{u},  \label{19}
\end{equation}

then a decoupled system of equations equivalent to the equations (\ref{7})
and (\ref{8}) is given by

\begin{equation}
\overset{\cdot }{\delta u_{1}}=0,  \label{20}
\end{equation}

\begin{equation}
\overset{\cdot }{\delta u_{2}}=\left( a+b\right) \delta u_{2}.  \label{21}
\end{equation}

The second of these equations can be written as

\begin{equation*}
\frac{d\delta u_{2}/dt}{\delta u_{2}}=\frac{d\ln \delta u_{2}}{dt}=a+b,
\end{equation*}

then $\delta u_{1}=C_{1}$ and $\delta u_{2}=C_{2}\exp \left[ \left(
a+b\right) t\right] .$ Multiplying vector $\delta \mathbf{u}$ by matrix $%
\mathbf{P,}$ we recover the old variables, resulting in

\begin{equation}
\delta n_{1}=C_{2}\frac{a}{b}\exp \left[ \left( a+b\right) t\right] +C_{1},
\label{22}
\end{equation}

\begin{equation}
\delta n_{2}=C_{2}\exp \left[ \left( a+b\right) t\right] -C_{1}.  \label{23}
\end{equation}

It is seen that $a+b$ is negative because it can be written as

\begin{equation}
a+b=-\left( p_{1}+p_{2}\right) -\left( q_{11}+q_{21}\right) n_{1}-\left(
q_{12}+q_{22}\right) \left( 1-n_{1}\right) ,  \label{24}
\end{equation}

but $0\leqslant n_{1}\leqslant 1$, then this sum is negative.

We can determine the constants appearing in Eqs. (\ref{22}) and (\ref{23}),
by using the initial condition, i.e. for $t=0$ we recover $\delta n_{1}(0)$
and $\delta n_{2}(0)$, then

\begin{equation}
\delta n_{1}\left( 0\right) =C_{2}\frac{a}{b}+C_{1},  \label{25}
\end{equation}

\begin{equation}
\delta n_{2}\left( 0\right) =C_{2}-C_{1}.  \label{26}
\end{equation}

From Eqs. (\ref{25}) and (\ref{26}) we can finally express the constants as
functions of the initial conditions, in the form

\begin{equation}
C_{1}=\frac{\delta n_{1}\left( 0\right) -\left( a/b\right) \delta
n_{2}\left( 0\right) }{1+a/b},  \label{27}
\end{equation}

\begin{equation}
C_{2}=\frac{\delta n_{1}\left( 0\right) +\delta n_{2}\left( 0\right) }{1+a/b}%
.  \label{28}
\end{equation}

Let's see what is the asymptotic behavior of $\delta n_{1}$ and $\delta
n_{2} $ given by Eqs. (\ref{22}) and (\ref{23}):

\begin{equation}
\underset{t\longrightarrow \infty }{\lim }\delta n_{1}=C_{1},  \label{29}
\end{equation}

\begin{equation}
\underset{t\longrightarrow \infty }{\lim }\delta n_{2}=-\text{ }C_{1},
\label{30}
\end{equation}

because as we saw, $a+b$ is negative.

Then, by adding the two equations above, we have

\begin{equation}
\underset{t\longrightarrow \infty }{\lim }\left( \delta n_{1}+\delta
n_{2}\right) =0.  \label{31}
\end{equation}

So this result means that a point slightly away from the straight line given
by $n_{2}=1-n_{1}$ would always tend to approach such line. In this sense
the line works as an attractor for the system. Besides, equations (\ref{29})
and (\ref{30}) tell us that upon a perturbation, the system does not return
to the original equilibrium point, but to a new point along the equilibrium
line, breaking the symmetry with respect to the line perpendicular to the
equilibrium line that passes through the perturbed point.

The condition that allows the perturbed system to return to the original
point is that $C_{1}=0$ (as inferred from Eqs. (\ref{29}) and (\ref{30})).
According to Eq. (\ref{27}) we see that:

\begin{equation}
C_{1}=0\Rightarrow b\text{ }\delta n_{1}\left( 0\right) =a\text{ }\delta
n_{2}\left( 0\right) .  \label{32}
\end{equation}

Substituting Eq. (\ref{32}) into (\ref{7}) and (\ref{8}) for $t=0$, we have

\begin{equation}
\overset{\cdot }{\delta n_{1}}\left( 0\right) \simeq \left( a+b\right)
\delta n_{1}\left( 0\right) ,  \label{33}
\end{equation}

\begin{equation}
\overset{\cdot }{\delta n_{2}}\left( 0\right) \simeq \left( a+b\right)
\delta n_{2}\left( 0\right) .  \label{34}
\end{equation}

So the equations become totally symmetric and consequently the perturbed
point undergoes a change that is fully equivalent in both directions.

Another interesting question we could ask ourselves is, what does the
direction of the breaking symmetry depend on? To answer this question,
imagine we make (without loss of generality) a displacement perpendicular to
the line of equilibrium, so that $\delta n_{1}\left( 0\right) =\delta
n_{2}\left( 0\right) .$ In this case we see that

\begin{equation}
C_{1}=\left( b-a\right) /\left( b+a\right) .  \label{35}
\end{equation}

When $b>a$, as we can see from Eqs. (\ref{29}) and (\ref{30}), the
equilibrium point will move to a bigger value of $n_{1}$ and, consequently,
a lower value of $n_{2}$ (and vice versa for $b<a$).

In the following calculations the result obtained above, i. e.  that the
equilibrium points correspond to the saturated market, will be re-obtained
for some particular cases. As we will see, the position of the equilibrium
point along the straight line of saturation will depend on the system's
coefficients. In the previous analysis the dependence on the coefficients
was via the constant $C_{1}$.

One lesson that remains from this section is that, for the proposed system,
there are no equilibrium points with a non null population of non-adopters.
This comes from the fact that all the influence coefficients are positive.
That is because the coefficients can be interpreted as probabilities (as is
seen in the original work by Bass \cite{Bass 69}). Also, the cross
coeficients $q_{ij}$ (with $i\neq j$) are positive too, because they
represent the observed positive influence of the adopters of one brand on
the potential adopters of the other brand. There is no observational support
for a negative cross influence. It is only in that case that we could find
equilibrium points that don't reach the market saturation.

\subsection{Dependence of the equilibrium points with the model parameters}

It is of great interest to determine the relationship between the model
parameters and the final equilibrium values. While this presents a difficult
analysis for the model as it is, the problem can be solved analytically for
the case without cross-brand influence. Therefore we will first analyze the
case where the cross-influence is negligible, and then compare that with the
cases in which this effect is taken into account. In both cases the standard
system of Bass-like equations, given by Eqs. (\ref{3}) and (\ref{4}) is used.

\subsubsection{Within-brand influence $(q_{12}=q_{21}=0)$}

This approach was introduced previously in ref. \cite{Kalish}, therein two
products competing in the same market are considered. Also in another
section of this work cross effects are considered, but in this case to model
the effect on the diffusion of the product in a market, when the same
product is introduced simultaneously in another market.

In this section we consider two brands competing for a common market,
neglecting the cross effect, i.e. in equations (\ref{3}) and (\ref{4}) we
neglect the corresponding coefficients. Then we have the following system of
equations:

\begin{equation}
\overset{\cdot }{n_{1}}\left( t\right) =\left( p_{1}+q_{11}n_{1}\right)
\left( 1-n_{1}-n_{2}\right) ,  \label{36}
\end{equation}

\begin{equation}
\overset{\cdot }{n_{2}}\left( t\right) =\left( p_{2}+q_{22}n_{2}\right)
\left( 1-n_{1}-n_{2}\right) .  \label{37}
\end{equation}

This system of equations can be solved analytically, so as to relate the
equilibrium point $(n_{1},n_{2})$ such that $n_{1}+n_{2}=1$, with the
coefficients of imitation and adoption of the model (see deduction on
Appendix A) , leading to:

\begin{equation}
1+\frac{q_{22}}{p_{2}}\left( 1-n_{1}\right) =\left( 1+\frac{q_{11}}{p_{1}}%
n_{1}\right) ^{\frac{q_{22}}{q_{11}}}.  \label{38}
\end{equation}

The equation above allows us to solve implicitly the dependence between the
proportion of adopters of brand 1 ($n_{1})$ and the coefficients $q_{11}$, $%
q_{22}$, $p_{1}$ and $p_{2}$. As for $n_{2}$, we can obtain it using $%
n_{2}=1-n_{1}$. In order to verify this result, for the particular case
where $q_{22}=q_{11}=q$ and $p_{1}=p_{2}=p$, Eq. (\ref{38}) gives $n_{1}=1/2$%
, as should be the case, since there would be no difference between the two
brands.

Some scenarios were developed solving Eq. (\ref{38}) and using the numerical
resolution method of Newton-Raphson. In Figure 1 $n_{1}$ and $n_{2}$ vs. $%
\delta q$ and $\delta p$ respectively are shown, for the following cases%
\begin{eqnarray*}
1)\text{ }p_{1} &=&p_{2}=0.03,q_{11}=0.2,q_{22}=q_{11}+\delta q,\text{ with }%
\delta q=0.,0.1,...,0.8. \\
\text{\ }2)\text{ }q_{22} &=&q_{11}=0.4,p_{1}=0.01,p_{2}=p_{1}+\delta p,%
\text{ with }\delta p=0.,0.01,...,0.09.
\end{eqnarray*}

Starting values {}{}were chosen near to the average values {}{}observed by
Sultan et al. (\cite{Sultan}) from to 213 real product cases, i.e. $%
<p>\simeq 0.03$ and $<q>\simeq 0.4.$

\begin{figure}
  \centering
    \includegraphics[width=0.9\textwidth]{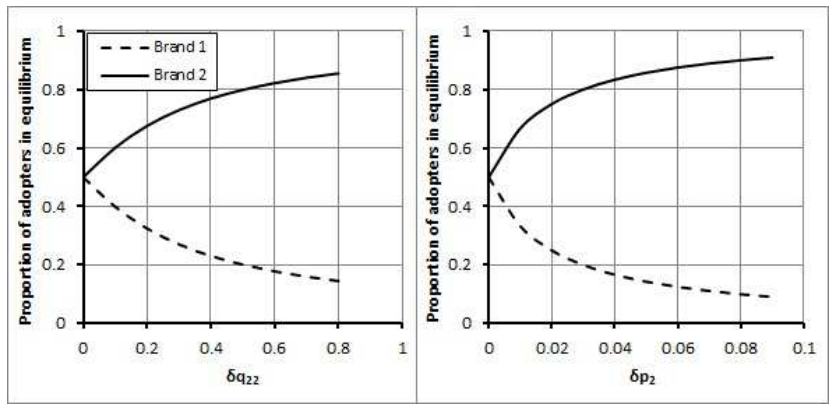}
  \caption{Proportion of adopters vs increments in the imitation and evaluation parameters respectively.}
  \label{fig1}
\end{figure}

On the left chart of Fig. (1) we can observe that an increase in the
coefficient of imitation strongly influences the final market proportion
reached, so that for a 250\% increase in $q_{22}$ (0.2 to 0.7), the final
market share for this brand is approximately 80\%. In the right chart it is
shown that the effect due to an increase in parameter $p$ is similar (in
percentage), where for a 250\% increase in $p$ (from 0.01 to 0.035) the
market share for brand 2 reaches 80\% as well.

\subsubsection{Cross-brand influence}

We analyze the influence of the cross-terms $q_{ij}$ $(i\neq j)$ in the
adoption dynamics given by the system of equations (\ref{3}) and (\ref{4}).
To do so, the system was solved numerically with the Runge-Kutta method. In
all calculations we considered the values ${}{}p_{1}=0.03$, $q_{1}=0.38$, $%
p_{2}=0.06$ and $q_{2}=0.68$, which are empirically reasonable according to
ref. (\cite{Sultan}). The values {}{}of the cross-terms were chosen in order
to make a systematic study, being then: A) $q_{12}=q_{21}=0,$ B) $q_{12}=0,$ 
$q_{21}=0.5,$ C) $q_{12}=0.5,$ $q_{21}=0$ and D) $q_{12}=0.5,$ $q_{21}=0.5$.
Corresponding diffusion curves are shown in Fig. 2.

\begin{figure}
  \centering
    \includegraphics[width=0.9\textwidth]{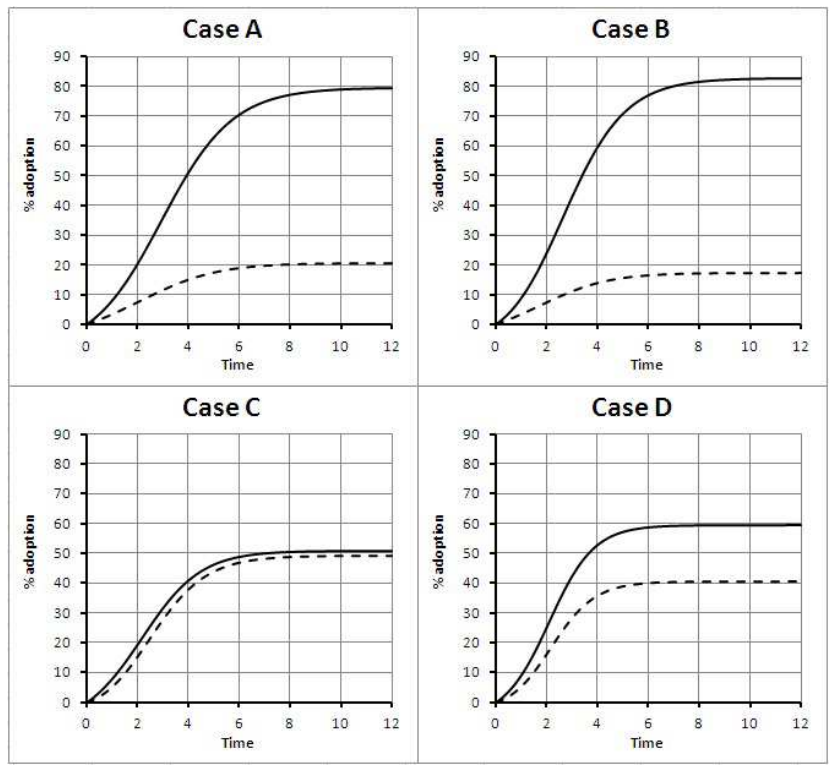}
  \caption{Study of the effect of cross-brand influence parameters. Brand 1 shown with dashed line and brand 2 with full line.}
  \label{fig2}
\end{figure}

Case "A" is a reference corresponding to when there not cross-brand
influence. In the case "B" by a non-zero value of $q_{21}$ the adoption of
the brand 2 is reinforced, as a result accentuates the difference in the
proportion of adopters of brand 2 compared to 1 in favor of the first.
However this effect is not linear, as expected, principally when the full
saturation with brand 2 is close. On the contrary by increasing the $q_{12}$
coefficient, such as we can appreciate in the case C, mainly tend to
approach, this effect being stronger than the previous case of widening. It
is interesting to note, that this effect would do compensate the advantage,
of brand 2 over the other, due to better personal evaluation of the utility,
making both brands finally have almost the same performance. Finally in the
case D where both cross coefficients has the same value, a compensation of
the effect produced in C occurs, i.e. some separation in the final
proportion it is observed.

\section{Agent-based model of innovation competition\protect\bigskip}

The application of the ABMs to the study of the diffusion of innovations is
currently a subject of great interest, as evidenced by the large amount of
research work in this area (see for example \cite{Olaru}, \cite{Schramm}, 
\cite{Chatterjee}, \cite{Goldenberg}). In these models the agents represent
individuals who choose between two options: to adopt or not the new product.

The immediate physical analogy comes from thinking the agent as if it were a
particle which can be in one of two states. As the system is constituted by
a large number of "particles" (agents), it is straightforward to think about
applying statistical mechanics to its study. A similar system is the
statistical Ising model, used in the field of physics. In this model the
agents are atoms arranged in a regular network with two possible spin
directions. The probability of finding an atom at a certain spin state can
be described in thermodynamic equilibrium terms by the Boltzmann-Gibbs
distribution \cite{Laciana-Rovere}. However, we shouldn't oversimplify by
not taking into account the additional capacities that the agents present in
real social systems, such as the learning possibility, the risk aversion,
and also the free will to determine his contacts. The proposed model is used
only as a formal frame that, due to its versatility, could be useful for the
future introduction of the characteristics mentioned above.

In the analog physical problem there is a local and an external effect. The
local effect is caused by the magnetic field produced by the neighborhood of
the considered particle and an the external effect is due to the external
field, which is not modified by the interaction with neighbors. In the
social problem the local effect is due to the tendency to imitate the
neighbors, and the external field is replaced by an individual assessment
(or utility perception) made by each agent according to the qualities of
product. However in the social context, the idea of neighborhood is somewhat
vague, since the regular network is not usually the most appropriate way to
describe the interaction.

A family of networks that are better suited for the description of social
interaction processes is the one known as \textquotedblleft small worlds
networks\textquotedblright\ (SWN) \cite{Watts}. Such networks may be
constructed from a regular network by a rewiring method introduced by Watts
and Strogatz \cite{Watts-Strogatz}. This method implies rewiring the
connections between agents by defining a parameter, called \textquotedblleft
probability of rewiring\textquotedblright , which is associated to each
connection, thus making possible for each connection between neighboring
agents to be broken and replaced with a connection between one of them to a
random agent on the network.

In \cite{Laciana et al.} there is a comparison between the diffusion curves
coming from the Bass macroscopic model and from the ABM with the Ising model
analogy. The impossibility of readoption is a characteristic of the Bass
model, and as such, it was introduced in the ABM for the comparison's sake.
In \cite{Laciana et al.} it is shown that for a given set of values of the
Bass model's parameters, the coincidence of the curves obtained from both
models is almost perfect, an example is shown in Figure 3:

\begin{figure}
  \centering
    \includegraphics[width=0.9\textwidth]{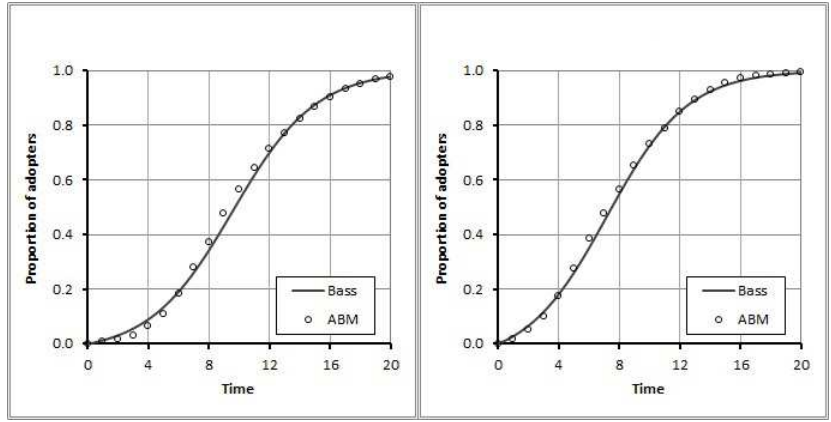}
  \caption{Example of the adoption curve match of an ABM and the Bass model.}
  \label{fig3}
\end{figure}

The adjustment of the curve obtained using the ABM, using the software
Mathematica 8 \cite{Mathematica}, was done by varying the parameters of the
Bass analytical curve, which describes the solution of equation (\ref{3})
when $q_{12}=0$, $p_{1}=p$, $q_{11}=q$, $n_{1}=n$ and $n_{2}=0$. That
solution is:

\begin{equation}
n\left( t\right) =\frac{1-e^{-\left( p+q\right) t}}{1+\left( q/p\right)
e^{-\left( p+q\right) t}}.  \label{39}
\end{equation}

\subsection{The three options algorithm}

In \cite{Laciana-Oteiza} we proposed a formalism that allows us to treat
problems that consider multiple options. It is based on an analogy of the
generalization for more than two options of the Ising model (known as the
Potts model). In the present article, this model is specialized for three
options (i.e.; \textquotedblleft brand 1\textquotedblright ,
\textquotedblleft brand 2\textquotedblright\ and \textquotedblleft
non-adoption\textquotedblright ). This allows us to propose the following
decision algorithm, which is the probability of finding an agent in a given
state (for null temperature). The example for state 1 is:

\ \ \ \ \ \ 
\begin{equation}
P\left( 1\right) =\left\{ 
\begin{array}{c}
1\text{ if }\Delta _{1k}\text{ }>\text{ }0\text{ for }k=2,3\text{ \ \ \ \ \
\ \ \ \ \ \ \ \ \ \ \ \ \ \ \ \ \ \ \ \ \ \ \ \ \ \ \ \ } \\ 
1/2\text{ if }\left\{ \Delta _{12}=0\wedge \Delta _{13}>0\right\} \vee
\left\{ \Delta _{13}=0\wedge \Delta _{12}>0\right\} \\ 
1/3\text{ if }\Delta _{12}=0\wedge \Delta _{13}=0\text{ \ \ \ \ \ \ \ \ \ \
\ \ \ \ \ \ \ \ \ \ \ \ \ \ \ \ \ \ \ \ \ \ } \\ 
0\text{ if }\Delta _{12}<0\vee \Delta _{13}<0\text{ \ \ \ \ \ \ \ \ \ \ \ \
\ \ \ \ \ \ \ \ \ \ \ \ \ \ \ \ \ \ \ \ \ }%
\end{array}%
\right. ,  \label{40}
\end{equation}

with

\begin{equation}
\Delta _{kj}=\nu _{k}-\nu _{j}+u_{k}-u_{j},  \label{41}
\end{equation}

for $k,j=1,2,3$.

The quantity $\nu _{j}$ is the proportion of agents in state $j$ within the
agent's contacts group, $u_{j}$ represents the utility of option $j$. The
quantity $\Delta _{kj}$ can be interpreted as an effective utility, in which
the first difference of Eq. (\ref{41}) introduces the imitation effect,
while the second difference compares the utilities of both.

\section{Comparison between Bass and ABM for two brands in competition}

In this section we will consider the adoption curves for two brands of an
innovative product in competition for a given market. We will compare the
results obtained by two very different models. In an way analogous to the
calculation done in ref. \cite{Laciana et al.}, we will consider the
adoption curve of an innovation obtained with an agent-based model
(micro-level), and the curve obtained from the Bass model (macro-level).

As highlighted in the work \cite{Libai} there is a dynamic influence on the
adoption of one of the brands by the interaction with the adopters of the
other brand, known as cross-brand effect. Mathematically this effect is
introduced by means of the parameter $q_{ij}$ , as we saw in equations (\ref%
{3}) and (\ref{4}). From the perspective of the agent based model the
cross-brand effect is not included. However it is interesting determine the
magnitude of the cross brand effect.

\subsection{Comparison for one brand}

As shown in ref. \cite{Laciana et al.}, for the case considering one brand
only, an almost perfect correspondence between the Bass model and the
agent-based model can be established (see Figure 3). The adjustment of the
Bass curve was made using Mathematica software \cite{Mathematica}. The
following data for the agent based model were used: utility difference
between adopting or not $\Delta u=0.6$, rewiring probability $p_{r}=0$
(regular lattice), early adopters dispersion $\sigma $ = uniform, rate of
incorporation of innovators (seeding) $\gamma =200$ per tick. The Bass curve
was obtained using equation (\ref{39}) with the parameters set to $p=0.0109$
and $q=0.3536.$These values {}{}are close to those determined by Bass \cite%
{Bass 94} (for the unmodified model) for air conditioners in the USA. These
are $p=0.0093$, $q=0.3798$, with errors of $0.0021$ and $0.0417$
respectively, which means the values {}{}obtained are empirically equivalent.

By increasing $\gamma $ until reaching $500$, which may be associated with a
more aggressive advertising campaign, and leaving the other variables
unchanged, we obtained the Bass parameter values {}{}$p=0.0239$ and $%
q=0.3513 $, which fits the result of ABM as shown in Figure 3 on the right
side.

\subsection{Comparison for two brands in competition}

The cases treated in isolation in the previous section will now be treated
as two brands in competition. These involve, besides parameters $p$ and $q,$
a parameter associated with the cross-brand effect, as discussed in Section
2, i.e. $q_{ij}\neq 0$.

In the following sub-section we will investigate what can be inferred from
the competition process between two brands. Our starting point will be our
previous knowledge of the separate diffusion curves for each of the two
brands in a monopolistic situation. The first case we will analyze is the
"minimal coincidence" one, in which the two models (Bass system and ABM)
reach the saturation point with the same proportion of adopters.

\subsubsection{Equilibrium proportions}

The first test will compare the results coming from the Bass model with the
ones coming from the ABM, particularly regarding the proportions of adopters
of brands $1$ and $2$ (i.e. $n_{1}$ and $n_{2}$) at the saturation point.
Let us consider two cases:

a) When we consider $q_{12}=q_{21}=0,$ the final proportion of adopters
reached by the Bass system is $n_{1}=0.3155$ and $n_{2}=0.68446$, while with
the ABM the result is $n_{1}=0.40125$ and $n_{2}=0.59875$.

b) For $q_{12}=q_{21}=0.077515$ we found that the coincidence with the ABM
result reaches five decimal places. The value obtained for the cross-brand
coefficients is within the range estimated in ref. \cite{Savin}, which may
vary from zero to values up to the order of magnitude of the direct
coefficients of the Bass system.

\subsubsection{Diffusion curves}

The following four numerical experiments were performed, which are shown in
Fig. 4:

1) In Fig. 4A we show the diffusion curves of two brands in competition with
the parameters $p$ and $q$ that we used before (in Fig. 3) to fit the
monopoly cases. These are $p_{1}=0.0109$, $q_{11}=0.3536$ for the lower
curve, and $p_{2}=0.0239$, $q_{22}=0.3513$ for the upper curve. In this case
only the within-brand effect is taken into account, i.e., $q_{12}=q_{21}=0$.
We see that the results for both models are very different.

2) Now we take into account the cross-brand effect, but restricted to $%
q_{12}=q_{21}$. We use the value that produces the best fit for the final
proportion of the adopters whic, as we saw in sub-section 4.2.1, is $%
q_{12}=q_{21}=0.0775$. All other parameters remain the same as in the last
case. The result is shown in Fig. 4B where there's an $8.04\%$ difference
between the areas under the curves.

3) As in the previows case, only the cross coefficients are modified, but
now without the imposed restriction. We use the values that better
approximate the curves coming from the ABM model wihtout changing the final
equilibrium. The result is shown in Fig. 4C. The difference between areas is
now $5.86\%$ and the values of the coefficients are $q_{12}=0.047$ and $%
q_{21}=0$.

4) In this experiment all the parameters are freely changed until reaching
an excellent fit between the curves, as we can appreciate in Fig. 4D. The
values obtained are: $p_{1}=0.0023$, $q_{11}=0.5787$, $p_{2}=0.0125$, $%
q_{22}=0.4252$, $q_{12}=0.0293$ and $q_{21}=0.0009$. The difference of area
obtained is $1.59\%$ and the fit coefficients for the curves 1 (the lower)
and 2 (the upper) respectively are: $R_{1}^{2}=0.9987$ and $R_{2}^{2}=0.9993$%
.

\begin{figure}
  \centering
    \includegraphics[width=0.9\textwidth]{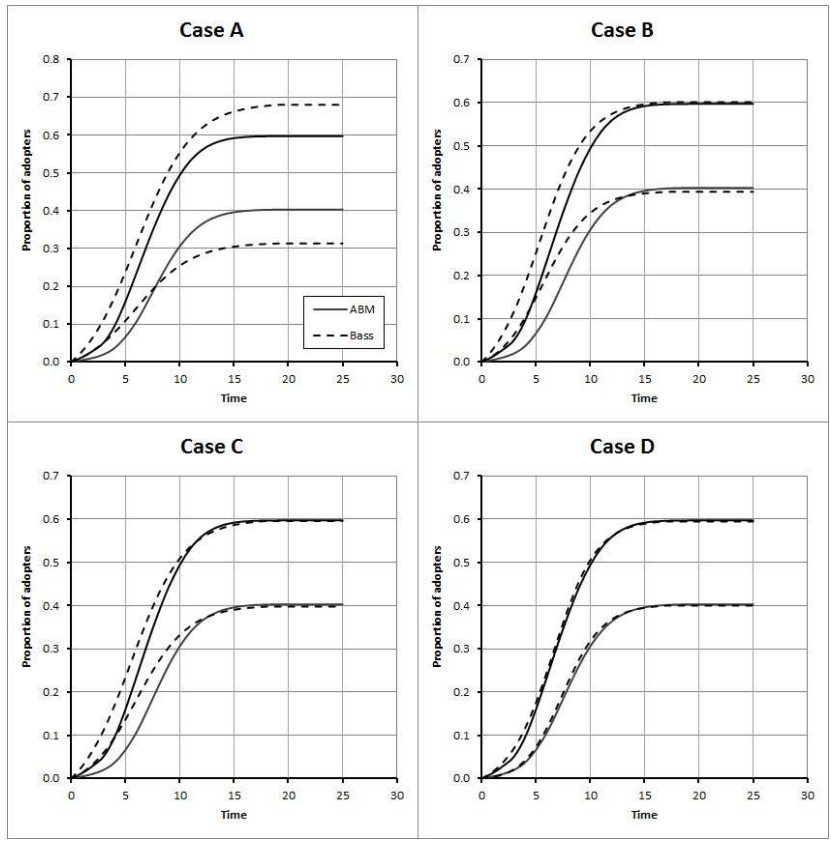}
  \caption{Comparisson between two diffusion models of two brands competing in a common market : AMB and Bass system, with different aproximations by the variation of cross-brand coefficients.}
  \label{fig4}
\end{figure}

\section{\protect\Large Conclusions and discussion}

As a general conclusion we have seen that the cross-brand coefficients of
the Bass system cannot be both zero if we want to fit the synthetic curves
obtained from an ABM inspired in the Potts statistical model.

In particular we highlight the following results:

\begin{itemize}
\item Assuming that all Bass system coefficients are positive, the only
equilibrium points are those that are on the straight line $n_{2}=1-n_{1}$
with $n_{1}>0,n_{2}>0$, which correspond to all the cases where the market
is saturated. The system is in indifferent equilibrium over this straight
line, in such a way that if we disturb the equilibrium and let the system
evolve freely, the result tends to return to the line. Hence, we can say the
line acts as an attractor.

\item For a minimal coupling of the system, that is, when only within-brand
influence is considered, it is possible to derive an analytical relationship
between the equilibrium points and the Bass coefficients. We see that both
personal evaluation and the rate of imitation strongly influence the market
share achieved at market saturation.

\item The effect of the cross-brand parameters is stronger when the
coefficient favors the brand with less utility (brand 1). So while a high
value of $q_{21}$ (such as $q_{21}=0.5$) increased the final $n_{2}$ in less
than $3\%$, a $q_{12}=0.5$ had a great influence, such that $n_{1}\sim n_{2}$
in the final equilibrium.

\item Regarding the comparison with the agent based model, we see that by
considering $q_{12}=q_{21}=0.077515$ the same result, in terms of the final
proportions, is obtained for both brands even though in the ABM there is
nothing comparable to the cross-brand terms of the Bass system equations.

\item The best fit for the diffusion curves obtained from both models is
reached when all the parameters of the Bass model are set free. We can
therefore conclude that the probabilities of innovation and imitation,
represented by $p$ and $q$ in the Bass model, depend on the competition
between brands. Therefore it is not correct to fix those coefficients a
priori, using only diffusion curves for the brands in a monopolistic market.
As a conclusion, we can say that cross-brand coefficients must be not null
for a proper fit.
\end{itemize}

These results leave some questions to be answered. In the first place we can
ask ourselves; why is it necessary to include the cross-brand terms $%
q_{_{jk}}\left( j\neq k\right) $, in the Bass system, in order to reproduce
the ABM curves?. As we have seen, it is not enough to consider the
parameters obtained from the monopolistic case to describe the duopolistic
one. Even when they provide a very good fit for the diffusion of both brands
as a monopoly, it is still necessary to include the cross-brand terms to
accurately model the diffusion of two brands in competition. This means that
in the ABM used there is something that models the cross brand effect, even
if it doesn't present any parameter equivalent to the cross-brand terms of
the Bass system. But, why?, does the ABM include a positive influence
towards the adoption of a brand via those agents that have adopted the other
brand?. The answer is yes. To realize this it is necessary to center in the
algorithm given in Eq. (40).

Let us think in a given agent that evaluates the possibility to adopt brand
1. The probability of adoption will be $P(1)$, given by Eq. (40). Let us
consider, for simplicity's sake, a didactic example in which both, products
1 and 2, have the same utility as the non-adoption option (state 3).
Considering this, Eq. (41) will be $\Delta _{kj}=\nu _{k}-\nu _{j}$. Let us
also consider that the agent is connected to 8 neighbours within a regular
network. We will compare two situations: (i) half the neighbours adopt brand
1 and the other half remains non-adopter and (ii) half the neighbours adopt
brand 1, a quarter of the neighbours adopt brand 2 and the remaining quarter
remains non-adopter. In case (i) we'll have $\Delta _{12}=1/2$ and $\Delta
_{13}=0$ which means $P(1)=1/2$, while case (ii) presents $\Delta _{12}=1/4$
and $\Delta _{13}=1/4$ which means $P(1)=1$.

In this example we can see that even if in both cases half the neighbourhood
has adopted brand 1, in the second case brand 1 has been promoted by the
agents that adopted brand 2. In other words we can say that there's an
influence in imitation at the category level, which seems reasonable
according to empirical data showing that the introduction of a new brand in
the market helps the diffusion existing analogous of the brand in the
context of innovations, as shown in ref.\cite{Libai}. This explains why it
is necessary to include the cross brand terms in the Bass system to compare
it with the ABM used.

Another question that arises from the results is; why is it not enough to
include the cross-brand terms to reach the best fit?.

We saw that using the best fit parameters for the monopolistic level plus
the cross-brand parameters isn't enough to achieve the best fit in the
duopolistic level. We need to free all of Bass parameters to do it.

This result is also explained by the kind of coupling between imitation and
cross-brand effect, which allows us to directly relate the macroscopic Bass
parameters to the microscopic ABM ones.

In the Bass system the $q_{_{ii}}$, related to the imitation probability,
are independent from the $q_{_{jk}}$. Nevertheless, in the ABM, as seen in
the previous example, the adoption of brand 2 boosts adoption of brand 1,
which implies that, in relative terms, the imitation effect seems stronger.
This dependence among the causers of adoption in the ABM means that, when
adjusting the Bass model, the coefficients $q_{_{ii}}$ also become dependent
of the $q_{_{jk}}$, which is why it is necessary to let them free to allow
for a mutual compensation in the best fit.

Future investigations could focus on relaxing some of the hypotheses used in
the present paper, in order to achieve a higher degree of realism. For
example, the conclusions we arrived to are subject to the supposition that
agents are homogeneous in their evaluation of both brands; we could
nevertheless suspect that the personal evaluation each person performs will
depend on his social status and that the interpersonal influences could have
a different weight depending on whether they are between agents of similar
social groups (homophily) or different social groups (heterophily) \cite{Li}%
. Recently, in ref. \cite{Bakshi} we can even find the possibility of
insertion of a product with an asymmetric attraction/repulsion, reflected by
a negative sign on one of the cross-brand coefficients. These interesting
ideas should be explored and will undoubtedly be subject of future
investigations.

\bigskip

\bigskip \pagebreak

\bigskip

\section{\protect\LARGE Appendix}

The purpose of this appendix is to establish relationships between the
proportions $n_{1}$ and $n_{2}$ and the parameters $p_{1}$, $p_{2}$, $q_{11}$
and $q_{22}$, when $q_{12}=q_{21}=0$.

We begin performing the quotient of Eqs. (\ref{36}) and (\ref{37}). After
that, we reorganize the terms in the form:

\begin{equation}
\frac{dn_{2}}{p_{2}+q_{22}n_{2}}=\frac{dn_{1}}{p_{1}+q_{11}n_{1}},  \tag{A1}
\end{equation}

by integrating we obtain:

\begin{equation}
\frac{1}{q_{22}}\ln \left( p_{2}+q_{22}n_{2}\right) =\frac{1}{q_{11}}\ln
\left( p_{1}+q_{11}n_{1}\right) +C.  \tag{A2}
\end{equation}

To determine the integration constant $C$ we will use the fact that Eq. (\ref%
{2}) holds at all times, in particular for the initial time, but we will
assume that the initial ratio is zero or $n_{1}(0)=n_{2}(0)=0$. Thus by
specializing of Eq. (A\ref{2}) at $t=0$ and solving for $C$ we get

\begin{equation}
C=\frac{1}{q_{22}}\ln p_{2}-\frac{1}{q_{11}}\ln p_{1},  \tag{A3}
\end{equation}

By replacing Eq. (A\ref{3}) in Eq. (A\ref{2}) and rearranging we have

\begin{equation}
\ln \left( 1+\frac{q_{22}}{p_{2}}n_{2}\right) =\ln \left[ \left( 1+\frac{%
q_{11}}{p_{1}}n_{1}\right) ^{\frac{q_{22}}{q_{11}}}\right] .  \tag{A4}
\end{equation}

Exponentiating Eq. (A\ref{4}) is

\begin{equation}
1+\frac{q_{22}}{p_{2}}n_{2}=\left( 1+\frac{q_{11}}{p_{1}}n_{1}\right) ^{%
\frac{q_{22}}{q_{11}}}.  \tag{A5}
\end{equation}

As Eq. (A\ref{5}) holds at all times, in particularly for the equilibrium,
that is when is valid the following equation

\begin{equation}
n_{2}=1-n_{1}.  \tag{A6}
\end{equation}

Substituting Eq. (A\ref{6}) in Eq. (A\ref{5}) yields Eq. (\ref{38}) which is
shown in subsection 2.2.1.

\begin{acknowledgement}
This research was supported partially by two U.S. National Science
Foundation (NSF) Coupled Natural and Human Systems grants (0410348 and
0709681) and by the University of Buenos Aires (UBACyT 00080).
\end{acknowledgement}

\end{document}